\documentstyle[preprint,aps]{revtex}

\newcommand{\be}{\begin{equation}}
\newcommand{\ee}{\end{equation}}
\newcommand{\bea}{\begin{eqnarray}}
\newcommand{\eea}{\end{eqnarray}}
\newcommand{\eq}[1]{Eq.~(\ref{#1})}

\begin{document}

%\preprint{\bf PREPRINT}
\columnsep0.1truecm
\draft

\title{Conserved Growth on Vicinal Surfaces}
\author{Harald Kallabis}
 
\address{
Center for Polymer Studies, Boston University, Boston, MA 02215, USA
}

\maketitle

\begin{center}
May 30, 1998
\end{center}

\begin{abstract}

A crystal surface which is miscut with respect to a high symmetry
plane exhibits steps with a characteristic distance.  It is argued
that the continuum description of growth on such a surface, when
desorption can be neglected, is given by the anisotropic version of
the conserved KPZ equation (T.~Sun, H.~Guo, and M.~Grant,
Phys.~Rev.~A~{\bf 40}, 6763 (1989)) with non-conserved noise.  A
one--loop dynamical renormalization group calculation yields the
values of the dynamical exponent and the roughness exponent which are
shown to be the same as in the isotropic case.  The results presented
here should apply in particular to growth under conditions which are
typical for molecular beam epitaxy.

\end{abstract}


\narrowtext

The fabrication of novel electronic devices requires experimental
conditions which are highly controllable. Molecular beam epitaxy (MBE)
is a very valuable technology for this purpose. In this process
particles are deposited under high vacuum conditions onto a crystal
surface which usually has been cleaved prior to growth.  The cleavage
process may generate a surface which is miscut against a
high--symmetry plane and which exhibits terraces with a characteristic
width. The terrace size can be made very large, but a small miscut can
never be avoided. Therefore the surface at the beginning of the growth
proces has always to be regarded as {\em vicinal}.

Since desorption of adatoms can be neglected under experimental
conditions which are typical for MBE, surface relaxation by adatom
diffusion is volume conserving.  In this paper I introduce the
continuum description of such {\em conserved} growth on a vicinal
surface. This allows for the analysis of the stochastic fluctations of
the surface, which appear on large time and length scales during
growth.  The surface fluctuations, apart from being of interest in
their own right, may be intimately related to the damping of
oscillations observed during layer--by--layer growth, as has been shown
recently \cite{Damping,DCSG}.

The surface fluctuations are expected to exhibit self--affine scaling
\cite{family_vicsek}:

\begin{equation}
w(t) \simeq a_{\perp}(\xi(t)/\tilde \ell \ )^{\zeta} 
\qquad {\mbox{and}} \qquad
\xi(t) \simeq \tilde \ell \ (t/\tilde t \ )^{1/z}
\label{scaling}
\end{equation}
if the surface is isotropic.  Here $w$ is the root mean square
variation of the film thickness (the surface width), $a_{\perp}$ the
thickness of one atomic layer, and $\xi$ is the correlation length up
to which the surface roughness has fully developed until time
$t$. $\zeta$ is the roughness exponent and $z$ the dynamical exponent.
The layer coherence length $\tilde \ell$ and the oscillation damping
time $\tilde t$ \cite{Damping} play the roles of natural cutoffs in
the continuum description of the surface fluctuations at small length
and time scales.  To calculate the values of $z$ and $\zeta$, we
derive the equation governing conserved growth on vicinal surfaces 
next.

On a coarse--grained scale the surface can be described at any given
time $t$ by a single--valued function $h(x_\parallel,x_\perp,t)$. The
coordinate system is chosen such that the surface tilt is $m$ in
$x_\parallel-$direction, while the steps are along the
$x_\perp-$direction.  It is then convenient to work in a tilted
coordinate system, $h\to h-m x_\parallel$, so that $h$ represents the
surface fluctuations around the average tilt.  Since we consider
conserved growth, the evolution equation for the surface has the form

\be
\label{start}
\partial_t h = -\nabla \cdot {\bf j} + \eta + {\cal F}
\ee
with a surface diffusion current ${\bf j}$ and a noise term $\eta$
which models the disorder entering the mesoscopic description
(\ref{start}) due to the stochastic nature of the growth process.
With the abbreviations $\partial_\parallel\equiv
\partial/\partial_\parallel$ and $\partial_\perp\equiv
\partial/\partial_\perp$, the vector $\nabla$ reads
$(\partial_\parallel,\partial_\perp)$.  ${\cal F}$ is the average
particle flux which is formally eliminated by changing to the comoving
frame, $h\to h-{\cal F}t$. All lattice constants are set to unity for
convenience.

The surface diffusion current has an equlibrium contribution and a
nonequilibrium contribution, ${\bf j}={\bf j}_{\rm eq}+{\bf j}_{\rm
neq}$.  ${\bf j}_{\rm eq}$ is given by the tendency of the adatom
current to even out gradients in the local equilibrium chemical
potential $n_{\rm eq}$ of the surface, ${\bf j}_{\rm eq}=-\Gamma\nabla
n_{\rm eq}$.  $\Gamma$ is the adatom mobility, which for simplicity is
assumed to be isotropic.  $n_{\rm eq}$ depends on $h$ like

\be
\label{GibbsThomson}
n_{\rm eq}=- \left(\frac{\kappa_\parallel}{\Gamma}\partial_\parallel^2
+\frac{\kappa_\perp}{\Gamma}\partial_\perp^2\right)h.
\ee
$\kappa_\parallel/\Gamma$ and $\kappa_\perp/\Gamma$ are the
anisotropic surface stiffnesses \cite{RSK,Dobbs,NozieresGodreche},
which for small variation of the surface can be regarded as constant.
\eq{GibbsThomson} represents the Gibbs--Thomson effect.

The nonequilibrium contribution ${\bf j}_{\rm neq}=-D\nabla n_{\rm
neq}$ to the surface current is driven by the nonequilibrium adatom
density $n_{\rm neq}$ \cite{villain91}. $D$ is the (isotropic)
diffusion constant.  On a flat surface without steps, $n_0$ is of the
order of $(F/D)\ell_D^2$, as derived in Ref.~\cite{villain92}.
$\ell_D$ denotes the typical island distance on a flat surface
\cite{WolfNATO}.  We assume $\ell_D$ to be isotropic, which is the
case if diffusion and lateral bonding of adatoms to islands is
isotropic.  When steps are present and the tilt is strong enough to
suppress island nucleation on terraces, i.e.\ if $|m|\ell_D\gtrsim 1$,
$n_0$ depends on the local surface tilt: $n_0\propto ({\cal
F}/D)/|m|^2$.  A convenient Ansatz for interpolation between those two
regimes is \cite{Pollain}

\be
n_{\rm neq}(\nabla h) = 
\frac{n_0}{1+\ell_D^2[(m+\partial_\parallel h)^2+(\partial_\perp h)^2]}
\ee 
in our coordinate system.  $n_{\rm neq}$ can be expanded for small
deviations from the global tilt to give

\be
\label{nneq}
n_{\rm neq}(\nabla h) \simeq n_{\rm neq}(0)
- \frac{\mu_\parallel}{D} \partial_\parallel h
- \frac{\lambda_\parallel}{D} (\partial_\parallel h)^2
- \frac{\lambda_\perp}{D} (\partial_\perp h)^2
\ee
with the quantities $n_{\rm neq}(0)$, $\mu_\parallel$, and
$\lambda_\perp$ being positive functions of $|m|$, $n_0$, and
$\ell_D$.  For small tilts, $|m|\ell_D\lesssim 1$, $\lambda_\parallel$
is positive, while for large tilts, $|m|\ell_D\gtrsim 1$, it is
negative.  These two cases distingush between nucleation--dominated
growth and step--flow growth, respectively.  The different signs of
the nonlinearities are known to have dramatic consequences for the
surface fluctuations in the case of nonconserved growth on vicinal
surfaces, which is described by the anisotropic KPZ equation
\cite{WolfVic}.  One aim of the present study is to see if a similar
scenario can be found in the conserved case.

The noise $\eta$ has three contributions in MBE growth
\cite{WolfNATO}.  Those are shot noise, diffusion noise, and
nucleation noise. Shot noise arises due to statistical fluctuations in
the atom beam which can be assumed to be isotropic. Diffusion noise
has its origin in the stochastic motion of adatoms. Since we assume
diffusion to be isotropic, this contribution is isotropic as
well. Finally, nucleation noise describes the random distribution of
island nucleation locations. Because diffusion noise and shot noise
together generate nucleation noise, the latter is also isotropic.  In
Ref.~\cite{Somfai} it has been shown that nucleation noise is
long--range correlated in time as long as the surface grows layerwise.
The continuum approach we pursue here is applicable for times larger
than the oscillation damping time, which marks the transition from
layer--by--layer growth to rough growth. After this time, the temporal
correlations have ceased.  Therefore we can assume nucleation noise to
be short--range correlated in time for the present purpose.  Note also
that despite their relation on the microscopic level, there are no
correlations between the different kinds of noise.  With these
remarks, the noise correlator reads \cite{WolfNATO,CKPZ}

\be
\label{ACKPZrealnoise}
\langle \eta(x,t)\eta(y,s) \rangle
=[ {\cal F} - {\cal D}\nabla^2 + {\cal N}(\nabla^2)^2]
 \delta^2(x-y) \delta (t-s).
\ee
${\cal F}$, ${\cal D}$, and ${\cal N}$ denote the strengths of
the shot noise, the diffusion noise, and the nucleation noise, 
respectively. The average value $\langle\eta\rangle$ vanishes.

In summary, we arrive at the anisotropic conserved KPZ (ACKPZ)
equation

\be
\label{ACKPZ}
\partial_t h=-\nabla^2[ 
(\kappa_\parallel\partial^2_\parallel + \kappa_\perp\partial^2_\perp)h
+ \mu \partial_\parallel h
+ \lambda_\parallel(\partial_\parallel h)^2
+ \lambda_\perp(\partial_\perp h)^2]
+\eta,
\ee
where the noise correlator is given by \eq{ACKPZrealnoise}.  Note that
the linear term proportional to $\mu$ cannot be transformed away as in
the nonconserved case \cite{WolfVic}.  This equation with $\mu=0$ has
first been studied in Ref.~\cite{Hove}. However, as the following
analysis shows, the conclusion presented there has to be corrected.
We study the surface fluctuations predicted by \eq{ACKPZ} next.

In the linear case $\lambda_\parallel=\lambda_\perp=0$, \eq{ACKPZ} can
be solved directly. One result is that the term $\propto\mu$ does not
influence the surface fluctuations.  Setting $\mu=0$, we may get the
values of the exponents by rescaling

\be
\label{Scale}
x_\perp\to bx_\perp, \qquad x_\parallel\to b^\chi x_\parallel
\qquad t\to b^z t, \qquad h\to b^\zeta h,
\ee
where $b$ is an arbitrary scaling factor \cite{Medina}.  The
anisotropy exponent $\chi$ \cite{WolfVic,Aniso} has to be introduced
here to account for the fact that in contrast to \eq{scaling}, which
is isotropic, there may be different characteristic lengths
$\xi_\parallel$ and $\xi_\perp$ governing the morphology of the
surface. $\chi$ is defined by the relation
$\xi_\parallel\propto\xi_\perp^\chi$.  By writing $b=\exp(d\ell)$ with
infintesimal $d\ell$, we get

\bea\label{kapar}
\frac{d\kappa_\parallel}{d\ell}&=&\kappa_\parallel(z-4\chi)\\
\label{rkappa}
\frac{d r_\kappa}{d\ell}&=&4 r_\kappa (1-\chi)\\
\label{eff}
\frac{d{\cal F}}{d\ell}&=&{\cal F}(z-2\zeta-2)\\\label{dee}
\frac{d{\cal D}}{d\ell}&=&{\cal D}(z-2\zeta-4)\\
\frac{d{\cal N}}{d\ell}&=&{\cal N}(z-2\zeta-6)\label{ennn}
\eea
for the change of the parameters in the continuum equation upon
rescaling, where $r_\kappa\equiv \kappa_\parallel/\kappa_\perp$.  From
Eqns.\ (\ref{eff}), (\ref{dee}), and (\ref{ennn}) we see that shot
noise is the most relevant type of noise, since it grows the fastest
upon an increase of scale for any values of $z$ and $\zeta$. For this
reason, we have to use \eq{eff} in the determination of the
exponents. Requiring scale invariance of the surface amounts to
setting the left--hand sides of the above equations to zero.  Using
Eqns.\ (\ref{kapar}), (\ref{rkappa}), and (\ref{eff}), we get the
exponents

\be
\label{linexp}
z=4, \qquad \zeta=1, \qquad \chi=1.
\ee
This means that a growing surface described by the linear version of
\eq{ACKPZ} can only be scale--invariant if the two spatial coordinates
are rescaled with the same scaling factor $b$. Then, the surface
fluctuations are governed by the isotropic version of \eq{ACKPZ} with
$\lambda_\parallel=0$ and $\lambda_\perp=0$.

The full (nonlinear) Eq.\ (\ref{ACKPZ}) is dealt with along the lines
described in Refs.\ \cite{Medina,FNS}: Wavenumbers
$b\pi/a\le|k_\perp|\le\pi/a$ and
$b^{\chi}\pi/a\le|k_\parallel|\le\pi/a$, where $a$ is the lattice
constant parallel to the surface, are integrated out ($a$ is set to
one for convenience). This is done in a one--loop approximation
published in detail elsewhere \cite{KallDiss}.  The resulting
renormalized parameters are then rescaled according to \eq{Scale}.  It
turns out that ${\cal F}$, $\lambda_\parallel$, and $\lambda_\perp$
are not renormalized.  The corresponding flow equations are

\bea\label{lapar}
\frac{d\lambda_\parallel}{d\ell}&=&\lambda_\parallel(z-4\chi+\zeta)\\
\label{rlambda}
\frac{d r_\lambda}{d\ell}&=&r_\lambda (1-\chi)
\eea
and \eq{eff}, where $r_\lambda\equiv \lambda_\parallel/\lambda_\perp$.
Those three equations already fix the exponents to be

\be
\label{nonlinexp}
z=\frac{10}{3} \qquad {\rm and} \qquad \zeta=\frac{2}{3}
\ee
if the system is scale invariant {\em and} if $\lambda_\parallel,
\lambda_\perp\ne0$.  These are the values for the isotropic conserved
KPZ equation in $d=2$ \cite{CKPZ,sgg} in one--loop order. (In two--loop
order they are slightly modified \cite{Janssen}.)

The one--loop corrections to the parameters do not depend on $\mu$
\cite{KallDiss}.  Thus the corresponding term does not play a role in
the determination of the surface fluctuations --- as in the linear
case.  (Interestingly this does not hold for a related deterministic
nonlinear equation which exhibits deterministic chaos \cite{Sato}.)
For this reason, we set $\mu=0$ in the following.

A remark on the noise renormalization is in order.  The nucleation
noise {\em is} renormalized and the corresponding flow equation reads

\be
\label{enn}
\frac{d{\cal N}}{d\ell}={\cal N}[z-2\zeta-6]
+g_\perp f(\kappa_\parallel,\kappa_\perp,\lambda_\parallel,
\lambda_\perp,{\cal F},{\cal D},{\cal N})
\ee
with $g_\perp\equiv (2\pi)^{-2}{\cal
F}\lambda_\perp^2/\kappa_\perp^3$.  $f$ is non--negative for
non--negative ${\cal F}$ or ${\cal D}$ \cite{KallDiss}.  This means
that, even if nucleation noise were absent $({\cal N}=0)$ initially,
this type of noise would automatically be generated by deposition and
diffusion noise --- which is immediately clear in the microscopic
picture. As mentioned above, however, we utilize \eq{eff} for the
determination of the exponents.

$\kappa_\parallel$ and $\kappa_\perp$ also are renormalized.  
This can be discussed most conveniently by considering the flow
equations for $r_\kappa$ and $g_\perp$:

\bea\nonumber
\frac{dr_\kappa}{d\ell}&=&\frac{g_\perp\pi}{4r_\kappa^{3/2}(r_\kappa-1)^2}
\left[
9(r_\lambda^2-r_\kappa^4)+r_\kappa^3(26-8r_\lambda)
\right.\\&& \label{rnu}
\left.+r_\kappa r_\lambda(8-26 r_\lambda) 
+16(r_\lambda-1)(r_\kappa^{5/2}+r_\kappa^{3/2}r_\lambda)
+r_\kappa^2(r_\lambda^2-1)
\right]\\
&&\nonumber\\
\frac{dg_\perp}{d\ell}&=&g_\perp\left[
2-\frac{3}{4}g_\perp\pi
\frac{
9r_\kappa^3+r_\kappa^2(7r_\lambda-26)-16r_\kappa^{3/2}(r_\lambda-1)
+r_\kappa(10r_\lambda+1)-r_\lambda
}{
r_\kappa^{3/2}(r_\kappa-1)^2
}
\right].\label{gperp}
\eea
Note that in the limit of $r_\lambda,r_\kappa\to1$, these flow
equations reduce to the isotropic case as obtained in
Ref.~\cite{CKPZ}.  Now the only important feature of these equations
is if the fixed point $g_\perp^*=0$ is stable.  If so, the
nonlinearities would vanish on large scales and the growth exponents
would take the values (\ref{linexp}) given by the linear equation.
This scenario is found for the anisotropic KPZ equation
\cite{WolfVic}.  Here we want to find out if the ACKPZ equation shows
the same behaviour.

For fixed $r_\kappa$, it is clear from \eq{gperp} that $g_\perp^*=0$
is unstable, since $dg_\perp/d\ell=2g_\perp$ for small $g_\perp$.
Since equations (\ref{rnu}) and (\ref{gperp}) are coupled, however,
$g_\perp^*=0$ could in principle be reached for $r_\lambda<0$ if
$r_\kappa$ vanishes together with $g_\perp$ in a suitable way, see
\eq{gperp}. To see if this is possible, we consider the limit
$r_\kappa\to0$ of (\ref{rnu}) and (\ref{gperp}),

\bea
\label{rkappasmall}
\frac{dr_\kappa}{d\ell}&=&\frac{9\pi g_\perp r_\lambda^2}{4r_\kappa^{3/2}}\\
\frac{dg_\perp}{d\ell}&=&g_\perp
\left[2+\frac{3\pi g_\perp r_\lambda}{4r_\kappa^{3/2}} \right].
\label{gperpsmall}
\eea
To simplify the notation we set $r_\lambda=-8/(3\pi)$.  For an initial
condition $g_\perp >r_\kappa^{3/2}$, the coupling constant $g_\perp$
will decrease initially, see \eq{gperpsmall}. For this decrease to
continue, $g_\perp/r_\kappa^{3/2}>1$ has to hold. \eq{rkappasmall}
then shows that $r_\kappa$, and consequently $g_\perp$ will increase
again.  Consequently, the fixed point $g_\perp^*=0$ is always
unstable.  Thus we can summarize that a change of universality class
from the nonlinear to the linear equation is not possible.

In conclusion I have argued that the ACKPZ equation describes the
fluctuations of a vicincal surface growing under MBE conditions.  The
dynamical and the roughness exponents are those of the isotropic
conserved KPZ equation.  In particular, the change from the nonlinear
to the linear universality class, as observed for the anisotropic KPZ
equation \cite{WolfVic}, is not found here.  Corroboration of these
findings in experiments or computer simulations are left for future
research.

{\em Acknowledgements.}
I thank Lothar Brendel and Martin Rost for very useful interactions
and a critical proofreading of the manuscript.  I am grateful to
Joakim Hove for making available his diploma thesis.  I acknowledge
support by the German Academic Exchange Service within the
Hochschulsonderprogramm III.

\end{document}